\documentstyle[12pt]{article}

\begin{document}
\topmargin 0pt
\oddsidemargin 7mm
\headheight 0pt
\topskip 0mm

\addtolength{\baselineskip}{0.40\baselineskip}

\hfill SNUTP 95-005

\hfill January, 1995

\hfill hep-th/9502078

\begin{center}
\vspace{0.5cm}
 {\large \bf ADM, BONDI MASS, AND ENERGY CONSERVATION IN
TWO-DIMENSIONAL DILATON
GRAVITIES }
\end{center}

\vspace{1cm}

\begin{center}
Won Tae Kim$^{(a)}$ and Julian Lee$^{(b)}$ \\
{\it Center for Theoretical Physics and Department of Physics, \\
Seoul National University, Seoul 151-742, Korea}
\end{center}

\vspace{1cm}

\begin{center}
{\bf ABSTRACT}
\end{center}

We show how a stress-energy pseudotensor can be constructed in two-dimensional
dilatonic gravity theories (classical, CGHS and RST) and derive the expression
for the ADM mass in these theories from it.
We define the Bondi mass for these theories by using the pseudotensor
formalism.
The resulting expression is the generalization of the expression for the ADM
mass. The boundary condition needed for the energy conservation is also
investigated. It is shown that under appropriate boundary conditions, our
definition of the Bondi mass is exactly the ADM mass minus the matter
radiation energy at null infinity.

\vspace{2cm}
\hrule
\vspace{0.5cm}
\hspace{-0.6cm}$^{(a)}$ E-mail address : wtkim@phyb.snu.ac.kr \\
\hspace{-0.6cm}$^{(b)}$ E-mail address : lee@phyb.snu.ac.kr \\
\newpage

\pagestyle{plain}

Recently there has been lots of interests in two-dimensional dilaton gravity
theories [1-5].  The main motivation comes from the fact that although
they
are
two-dimensional toy models, they possess most of the interesting properties of
the
four-dimensional gravity theories such as the existence of the black hole
solutions, Hawking radiations, etc. At the same time, they are more amenable to
quantum treatments than their four-dimensional counterparts. Thus, it is hoped
that they will pave a way to address interesting issues in quantum gravity such
as information loss problem [6] in black hole physics.

It is a well known fact that one cannot construct a conserved stress-energy
tensor
in general relativity except for space-times having particular
symmetries [7].
The fact that the stress-energy tensor for the matter fields alone is not
conserved
is not surprising since they exchange energies and momenta with the
gravitational field.
Furthermore, there is no notion corresponding to the stress-energy of the
gravitational field which is a generally covariant tensor. However, we can
introduce the concept of stress-energy for the
gravitational field if we take the view that the general relativity can be
treated
as a
spin-2 field theory in Minkowski background. The stress-energy so constructed
will be a pseudotensor, in the sense they are not generally covariant, but is
Lorentz covariant with respect to the background Minkowski metric.
Just as in the case of four-dimensional Einstein gravity, we can show that the
pseudotensor corresponding to energy density is a total derivative for the
two-dimensional dilaton gravity theories. Therefore for a asymptotically flat
space-time, the energy
becomes a surface term defined at either spatial or null infinity. The former
is the Arnowitt-Deser-Misner(ADM) mass [8] and the latter is called the Bondi
mass [9]. The stress-energy pseudotensor is constructed in such a way that it
is a conserved
current.  Then it is obvious that the difference of Bondi and ADM mass is
integral of this current flowing out to null infinity. In the four-dimensional
Einstein gravity,
this current is interpreted as the flux of radiation energy. (It is to be
understood that the word radiation refers to massless fields.)
In the case of two-dimensional dilaton gravity theories, the graviton and
dilaton fields have no
propagating degrees of freedom and only the matter radiation is capable of
escaping to null infinity.

  In this paper we consider the energy conservation in two-dimensional dilaton
gravity theories, which are classical,
Callan-Giddings-Harvey-Strominger(CGHS) [1] and
Russo-Susskind-Thorlacius(RST) [2,3] models. We construct a
stress-energy pseudotensor for these theories and rederive the well known
expression for the ADM mass. We also specify the asymptotic boundary
conditions needed for the total energy conservation.  We then define the notion
of Bondi mass by integrating
the stress-energy pseudotensor along the null
hypersurface instead of spatial one. The resulting expression is the
generalization of the known expression for the ADM mass.  It will be shown
that our definition of Bondi mass is ``correct" in the
sense that it is exactly the ADM mass minus the integral of radiation
energy flux at
null
infinity under reasonable boundary conditions. This result can be considered as
 the energy conservation applied to ADM and Bondi mass.

We consider dilaton gravity theories described by the action,
\begin{eqnarray}
S \hspace{-0.2cm}&=& \hspace{-0.2cm} \frac{1}{2\pi} \int d^2 x \sqrt{-g}\,
\big[(e^{-2\phi}-\frac{\kappa}{2}\phi) R + 4 e^{-2\phi}( (\nabla
\phi)^2 +\lambda^2) -\frac{1}{2} Q^2 R \frac{1}{\nabla^2 } R \nonumber \\
&\qquad& \qquad \qquad \qquad - {1 \over 2}
  \sum^{N}_{i=1} (\nabla
f_i)^2 \big]
\label{dil}
\end{eqnarray}
where $\kappa = Q = 0$ corresponds to the classical action, $\kappa =0,\ 2Q^2 =
{N \over 12}$
gives the CGHS model [1], and $\kappa = 2Q^2 =  {N-24 \over 12} $ corresponds
to
RST model [2,3]. The latter two are the one-loop effective action incorporating
the
effect of trace anomaly of matter fields.

  It is convenient to split the action above as
\begin{equation}
S = S_{DG} + S_{matter}
\end{equation}
where
\begin{eqnarray}
S_{DG} &\equiv& {1 \over 2\pi} \int d^2 x \sqrt{-g}\, e^{-2\phi}
\left[ R + 4  (\nabla
\phi)^2 + 4 \lambda^2  \right], \nonumber \\
S_{matter} &\equiv& {1 \over 2\pi} \int d^2 x \sqrt{-g}\,
\left[ - {1 \over 2} \sum^{N}_{i=1} (\nabla
f_i)^2  -{\kappa \over 2}\phi R  -{1 \over 2} Q^2 R {1 \over \nabla^2 } R
\right].
\end{eqnarray}
$S_{DG}$ governs the dynamics of the dilaton-gravity fields which is treated
classically, whereas $S_{matter}$
governs the classical dynamics of the scalar fields $f_i$ and their possible
one-loop effects in the large $N$ limit [1,2,3].
  Next, one expands the graviton-dilaton
fields around the linear dilaton vacuum(LDV),
\begin{equation}
g_{\mu \nu} = \eta_{\mu \nu} + h_{\mu \nu}, \quad \phi = -\lambda x^\alpha
\eta_{
\alpha \beta} \epsilon^\beta + \psi
\end{equation}
where $\eta_{11} = -\eta_{00}=1, \eta_{01} = \eta_{10}=0$ and $\epsilon^\alpha$
satisfy
$\epsilon^\alpha \eta_{ \alpha \beta} \epsilon^\beta = 1$. Note that the
Poincar\'e symmetry present in the original Lagrangian is spontaneously broken
by
the vector $\epsilon^\alpha$. We now have a preferred coordinates
$(x^0=t,x^1=q)$ such
that  $x^\alpha \eta_{ \alpha \beta}
\epsilon^\beta
= q$, which we will take from now on. Therefore, in contrast to the case of
four-dimensional
Einstein gravity where one has the
energy-momentum four-vector for the whole system, only the concept of energy is
meaningful in the case of two-dimensional dilaton gravities, which is usually
called mass.
Another consequence of this broken
symmetry is the fact that the stress-energy pseudotensor constructed by the
generalized Belinfante procedure [10] does not give any reasonable expression
for the ADM mass [11] when integrated over a spatial hypersurface
approaching spatial infinity.

 To construct the stress-energy pseudotensor, one linearizes the equation of
motion for the
dilaton-graviton fields given by
\begin{equation}
G_{ \mu \nu} =  T_{\mu \nu} \label{eom}
\end{equation}
to get [12]
\begin{equation}
G^{(1)}_{ \mu \nu} =  T_{\mu \nu} - G^{(2)}_{\mu \nu}  \label{linear}
\end{equation}
where
\begin{eqnarray}
G_{ \mu \nu} &\equiv& 2 \pi {1 \over \sqrt{-g}} { \delta S_{DG} \over \delta
g^{\mu \nu} } = 2e^{-2\phi}\left[ \nabla_\mu  \nabla_\nu \phi + g_{\mu \nu} (
(\nabla \phi )^2 -
\nabla^2 \phi -\lambda^2 ) \right], \nonumber \\
T_{ \mu \nu }  &\equiv& -2 \pi {1 \over \sqrt{-g}} { \delta S_{matter} \over
\delta g^{\mu \nu} },
\end{eqnarray}
and $G^{(1)}_{\mu \nu}$ is the part of $G_{\mu \nu}$ linear in $h_{\mu
\nu}$ and $\psi$, and $G^{(2)}_{\mu \nu}$ is the part which is of second or
higher
order. Then
the second term on the right-hand side of
Eq.\,(\ref{linear}) can be
interpreted as a stress-energy of the dilaton-graviton fields. The energy
current,
\begin{equation}
j^\mu \equiv T^{\mu \nu} - G^{(2) \mu \nu}
\end{equation}
satisfies the conservation law, where
it is to be understood that the Minkowski metric is used in the lowering and
uppering of indices. To see this, we
note that $G^{(1) \mu 0} \equiv \eta^{\mu \alpha} \eta^{0 \beta}
G^{(1)}_{\alpha
\beta}$  is in the form which is identically conserved. This fact follows from
the so called linearized Bianchi
identity, which in turn follows from the gauge invariance.
The gauge transformation corresponding to a diffeomorphism is given by
\begin{eqnarray}
 \delta g^{\mu \nu} &=& \epsilon^\alpha \partial_\alpha g^{\mu\nu}
-\left(\partial_\alpha \epsilon^\mu\right) g^{\alpha \nu} -
 \left(\partial_\alpha \epsilon^\nu \right) g^{\mu \alpha }, \nonumber \\
\delta \phi &=& \epsilon^\lambda \partial_\lambda \phi, \nonumber \\
\delta f_i &=& \epsilon^\lambda \partial_\lambda f_i  \label{gtr}
\end{eqnarray}
where $\epsilon(x)$ is an arbitrary function parameterizing the gauge
transformation. We note that not only the whole action is invariant under the
transformation Eq.\,(\ref{gtr}), but also the dilaton-gravity part of the
action
$S_{DG}$ is gauge invariant. From this fact we get the Bianchi identity:
\begin{equation}
 (\partial_\lambda g^{\mu\nu}
) {\delta S_{DG} \over \delta g^{\mu \nu }}+ 2 \partial_\alpha \left( g^{\alpha
\nu}
{\delta S_{DG} \over \delta g^{\lambda \nu }} \right)
 + \left( \partial_\lambda \phi \right) {\delta S_{DG} \over \delta \phi }   =
0.
\label{id}
 \end{equation}

Since Eq.\,(\ref{id}) is an identity, it will hold order by order when we
expand
it in terms
of $h_{\mu \nu}$ and $\psi$. To linear order, one has
\begin{equation}
2 \partial_\alpha  \eta^{\alpha
\nu}
G^{(1)}_{\lambda \nu}
 -\lambda \eta_{\lambda \alpha} \epsilon^\alpha  F^{(1)} = 0 \label{lid}
 \end{equation}
 where $F^{(1)}$ are the parts linear in
$h_{\mu \nu}$ and $\psi$ of   ${\delta S \over \delta \phi}$.
Eq.\,(\ref{lid}) is called the linearized Bianchi identity.
In the
$(t,q)$ coordinate given above, the linearized Bianchi identity (\ref{lid})
gives
\begin{equation}
\partial_\mu G^{(1) \mu 0} \equiv 0. \label{lc}
\end{equation}
Showing that $G^{(1) \mu 0}$  is a conserved current.
Therefore we define $G^{(1) \mu \nu}$, or alternatively $T^{(1) \mu \nu}-G^{(2)
\mu \nu}$, as the stress-energy pseudotensor.

 Although $j^\mu$ satisfies the local
conservation law due to Eq.\,(\ref{linear}) and Eq.\,(\ref{lc}), it does not
follow immediately that the total energy
 \begin{equation}
E_{ADM} \equiv  \int_{-\infty}^\infty dq  j^0 = \int_{-\infty}^\infty dq
G^{(1)
0 0}  \label{admcur}
\end{equation}
 is conserved without specifying the boundary conditions. However, if we
consider a space-time which approaches the LDV at spatial infinity fast enough,
then
\begin{equation}
j^1 =  T^{1 0}  - G^{(2) 1 0} \label{outflow}
\end{equation}
would vanish at $q \to \pm\infty$ and the total energy
$E_{ADM}$ becomes time-independent.
 On the other hand   $G^{(1) \mu 1}$, which would correspond to the momentum
current, does not even satisfy the local conservation law
since the translational symmetry in spatial
direction is spontaneously broken by the LDV.
Using the expression at the end of Eq.\,(\ref{admcur}), one gets after some
straightforward algebra
 \begin{equation}
  E_{ADM} = 2 e^{2 \lambda q} (\partial_q \psi + \lambda {h_{11} \over
2})\vert_{q\to \infty}.
\label{old}
  \end{equation}
  It is easy
to see that the contribution from $q \to -\infty$
vanishes.  Writing $g_{11}=e^{2\rho}$, we get the expression
\begin{equation}
E_{ADM} = 2 e^{2 \lambda q} (\partial_q\psi + \lambda \rho ) \vert_{q\to
\infty}
\label{conf}
\end{equation}
if $\rho$ falls to zero at $q \to \infty$ not slower than $e^{-2\lambda q}$.
Eq.\,(\ref{conf}) is the expression
used often
in
the literature [13,14]. Although the conformal gauge is commonly used in
order to get this expression, note that we only imposed the condition that the
dilaton-graviton fields should approach LDV fast enough so that the current
$j^1$ given by
Eq.\,(\ref{outflow}) should vanish as $q \to \pm \infty$. We see that as long
as
the dilaton-gravity fields approach LDV configuration and the classical matter
fields $f_i$
vanish at $q \to \pm \infty$ the first term $T^{10}$ as a whole vanishes.
Also, since $G^{(2) \mu \nu}$ consists of terms quadratic or higher order in
$\rho,\psi$ times the common factor $e^{-2\phi}$ which behaves as $e^{2\lambda
q}$ at $q \to \pm \infty$, we easily see that it vanishes as $q \to -\infty$.
The
only nontrivial condition is given for $q\to\infty$, since the exponential
factor
blows up. It is sufficient to require that
\begin{equation}
 h_{\mu \nu} \sim e^{-2\lambda q}, \quad \psi \sim
e^{-2\lambda q} \label{oldbc}
\end{equation}
as $q \to \infty$. Then the expression (\ref{old}), or equivalently
(\ref{conf}), gives the
ADM mass which is time-independent. (Naively one might be tempted just to
impose the condition $e^{\lambda q} h_{\mu \nu} \to 0, \ e^{\lambda q} \psi \to
0$ at $q \to \infty$ so that $G^{(2) \mu \nu}$ vanishes. However, this
condition would not be interesting since if $\rho$ and $\psi$ vanish faster
than
$e^{-\lambda q}$ but slower than $e^{-2\lambda q}$ this configuration is a
physically pathological case since the ADM mass blows up.)

 Now let us introduce the light-front coordinates $\sigma_\pm \equiv t \pm q$.
The Bondi mass $B(\sigma^-)$ can then be defined in a straightforward manner
by
integrating the energy flux along the null line $t-q=\sigma^-$, {\it i.e.}
\begin{eqnarray}
B(\sigma^-) &\equiv& {1 \over 2} \int_{-\infty}^\infty d\sigma^+ j^-(\sigma^+,
\sigma^-)
\nonumber \\
&=& {1 \over 2} \int_{-\infty}^\infty d\sigma^+ G^{(1) - 0}(\sigma^+, \sigma^-)
\nonumber \\
&=& 2e^{2 \lambda q}  (\partial_q\psi+\lambda \rho) \vert_{\sigma^+ \to
\infty} \nonumber \\
&=& 2e^{\lambda (\sigma^+ -\sigma^-)}  ((\partial_+ - \partial_-)\psi+\lambda
\rho) \vert_{\sigma^+ \to
\infty}  \label{bondi}
\end{eqnarray}
where $G^{(1) - 0} \equiv G^{(1) 00}-G^{(1) 10}$ and the second line of
Eq.\,(\ref{bondi}) follows from Eq.\,(\ref{linear}).
Note that we have $$\lim_{\sigma^- \to -\infty} B(\sigma^-) = E_{ADM}.$$

Next, we will investigate the boundary condition required at the null
infinity so that one can interpret $B(\sigma^-)$ as the energy left in the
system
after the radiation has been emitted. Before that, we will compare the results
of this section with
other expressions of ADM and Bondi mass found in the literature and show they
agree. The following expression for the ADM mass
\begin{equation}
E_{ADM} = 2e^{2 \lambda q} ( \partial_q + \lambda) ( \psi - \psi^2
)_{q\to\infty} \label{new}
\end{equation}
was obtained by Bilal and Kogan [15] using the Hamiltonian formalism. They
required the boundary conditions at spatial infinity,
\begin{eqnarray}
 \psi \sim e^{-\lambda q}, &\quad& \rho \sim e^{-\lambda q}, \nonumber \\
e^{2 \lambda q} ( \psi - \rho) &\to& 0. \label{newbc}
\end{eqnarray}
 By imposing the conditions (\ref{oldbc}) and (\ref{newbc}) simultaneously,
{\it
i.e.} requiring
\begin{eqnarray}
 \psi \sim e^{-2\lambda q}, &\quad& \rho \sim e^{-2\lambda q}, \nonumber \\
e^{2 \lambda q} ( \psi - \rho) &\to& 0 , \label{newnewbc}
\end{eqnarray}
one can replace $\rho$ in Eq.\,(\ref{conf}) by $\psi$ and drop the $\psi^2$ in
Eq.\,(\ref{new}), giving the common expression
\begin{equation}
E_{ADM} = 2e^{2 \lambda q} ( \partial_q + \lambda) \psi |_{q\to\infty}.
\label{newnew}
\end{equation}

A different form of stress-energy pseudotensor was obtained in Ref.\,[11] as a
Noether current associated with the diffeomorphism invariance of the dilaton
gravity Lagrangian. The ADM mass obtained from this pseudotensor is
\begin{equation}
E_{ADM} = -{1 \over 2}{1 \over \sqrt{-g}}\epsilon^{\mu\nu}\left[ e^{-2 \phi}
\nabla_{\mu} e_{\nu} +{1 \over 2} e_{\mu} f_i \nabla_{\nu} f_i \right]
\end{equation}
where $e^0 = 1$ and $e^1=0$, and the uppering and lowering of indices are done
by
using the full metric $g_{\alpha\beta}$. Expanding the metric and dilaton
fields
around the LDV configuration, we again get the expression which agrees with
Eq.\,(\ref{conf}) under the boundary condition
\begin{equation}
e^{2 \lambda q} h_{10} \to 0
\end{equation}
together with Eq.\,(\ref{oldbc}).

Also we note that in case of CGHS model, a different formula for Bondi mass
was obtained in Ref.\,[1],
\begin{equation}B(\sigma^-) = 2e^{\lambda (\sigma^+ -\sigma^-)}  ((\partial_+ -
\partial_-)\psi+\lambda
\rho)  + {N \over 12} (\partial_- - \partial_+) \rho \vert_{\sigma^+ \to
\infty}, \label{newbondi}
\end{equation}
which is supposed to hold for general conformal coordinates.
As was noted in Ref.\,[1], the second term vanishes in the coordinates where
the dilaton-gravity fields approach the standard form of LDV  $\rho = 0, \phi =
-{\lambda \over 2} (\sigma^+ - \sigma^-)$ at $\sigma^+ \to \infty$, and in this
case
Eq.\,(\ref{newbondi}) reduces to Eq.\,(\ref{bondi}).

Now we consider the difference between $E_{ADM}$ and $B(\sigma^-)$ and
investigate the asymptotic boundary conditions under which it equals the matter
radiation energy emitted to the null infinity. Obviously, it
can be
represented by
the integral of the current flux along the null line at infinity from the point
$(\sigma^+,\sigma^-) =(\infty,-\infty)$ to the point
$(\infty, \sigma^-)$,
\begin{eqnarray}
  E_{ADM}-B(\sigma^-) \!\!&=&\!\! \lim_{ \sigma^+ \to \infty} {1 \over 2}
\int_{-\infty}^{\sigma^-}
d\sigma^- j^+ (\sigma^+,\sigma^-) \nonumber \\
\!\!&=&\!\! \lim_{ \sigma^+ \to \infty} {1 \over 2}
\int_{-\infty}^{\sigma^-}
d\sigma^- G^{(1)+0}(\sigma^+,\sigma^-) \nonumber \\
\!\!&=&\!\! -\!\!\lim_{ \sigma^+ \to \infty}
\int_{-\infty}^{\sigma^-}
d \sigma^- \partial_-\left[ 2e^{ \lambda (\sigma^+ -
\sigma^-) } \left( (\partial_+ -
\partial_+) \psi + \lambda \rho \right)(\sigma^+, \sigma^-)
\right]   \label{bch}
\end{eqnarray}
where the second line follows from Eq.\,(\ref{linear}), as before.
Again, we
want to find the boundary condition such that the contribution from
$G^{(2) \mu \nu}$ to $j^\mu$ flux is negligible.  We see that it is sufficient
to require
\begin{eqnarray}
\psi \sim e^{-\lambda \sigma^+}, &\quad& \rho \sim e^{-\lambda \sigma^+},
\nonumber \\
 e^{\lambda \sigma^+} h_{\pm\pm} \to 0,
    \label{nulbc}
\end{eqnarray}
as $\sigma^+ \to \infty$. Again, at $\sigma^+ \to -\infty$ it is enough to
require
that the configuration
approaches the LDV.
Indeed, for this boundary condition the total matter radiation energy flux at
null infinity is given by
\begin{eqnarray}
 &&\int_{-\infty}^{\sigma^-} d\sigma^- \sqrt{-g}\, T^{+0}|_{
\sigma^+ \to \infty} \nonumber \\
 &\equiv&  {1 \over 2} \int_{-\infty}^{\sigma^-}
d\sigma^- \sqrt{-g}\,(T^{++} + T^{+-})|_{\sigma^+ \to \infty} \nonumber \\
&=& \int_{-\infty}^{\sigma^-} d\sigma^- e^{-2(\rho + \phi)} ( 2
\partial_-^2 \phi - 4 \partial_- \rho
\partial_- \phi - 2 \partial_+ \partial_- \phi + 4 \partial_+ \phi
\partial_- \phi + \lambda^2 e^{2 \rho} )|_{\sigma^+ \to \infty} \nonumber \\
&=& \int_{-\infty}^{\sigma^-}
d\sigma^- \partial_-\left[ 2e^{ \lambda (\sigma^+ - \sigma^-) } \left(
(\partial_+ -
\partial_-) \psi + \lambda \rho \right)
\right]_{\sigma^+ \to \infty} \label{nulflux}
\end{eqnarray}
where the third line follows from the equation of motion (\ref{eom}).
Comparing Eq.\,(\ref{bch}) with Eq.\,(\ref{nulflux}), we see that
\begin{equation}
E_{ADM}-B(\sigma^-) = \lim_{\sigma^+ \to \infty} \int_{-\infty}^{\sigma^-}
d\sigma^- \sqrt{-g(\sigma^+, \sigma^-)}\,
T^{+0}(\sigma^+,\sigma^-).
\end{equation}
  Thus, we have shown that under the boundary condition (\ref{nulbc}), our
definition
of Bondi
mass agrees with the notion that the Bondi mass is the energy left in the
system
after the radiation has been emitted to null infinity.

In this paper, we showed how the expression for the ADM mass can be
derived using a stress-energy pseudotensor. We then defined the notion of Bondi
mass which is consistent with the
usual definition as the energy left in the system after
the radiation has been emitted. We also investigated the boundary
conditions required at spatial and null infinities in order for the energy
conservation to hold. As in the case of four-dimensional Einstein gravity,
these
asymptotic
boundary conditions were crucial in establishing the energy conservation.
The fact that the energy conservation is violated when these boundary
conditions
are not satisfied simply tells us that we cannot treat the system as being
``isolated" one
living on the Minkowski background anymore. The concept of energy is
meaningless
in this case.

 Another attempt of constructing Bondi mass
can be found in Ref.\,[16]. However, we are puzzled by the fact it violates the
energy
conservation even for reasonable boundary conditions such as in the case of a
evaporating black hole in RST model. On the other hand, it
is claimed in Ref.\,[16] that the Bondi mass is positive definite, whereas our
Bondi mass is not
necessarily so. We believe our definition of Bondi mass is more sensible since
it satisfies the energy conservation. By applying our definition of the Bondi
mass to the
case of the evaporating black hole in RST model, we get the correct energy
conservation in variance with the results of Ref.\,[16].  The details can be
found in Ref.\,[17].

\noindent
{\bf Acknowledgments} \break
We were supported by the Korea Science
and Engineering Foundation through the Center for Theoretical Physics (1995).

\end{document}